\documentclass[twocolumn,english,aps,english,preprintnumbers,groupedaddress,amsmath,amssymb,pra]{revtex4}

\usepackage[T1]{fontenc}
\usepackage[latin9]{inputenc}
\usepackage{amsmath}
\usepackage{amssymb}
\usepackage{graphicx}
\usepackage{esint}

\makeatletter
\@ifundefined{textcolor}{}
{%
 \definecolor{BLACK}{gray}{0}
 \definecolor{WHITE}{gray}{1}
 \definecolor{RED}{rgb}{1,0,0}
 \definecolor{GREEN}{rgb}{0,1,0}
 \definecolor{BLUE}{rgb}{0,0,1}
 \definecolor{CYAN}{cmyk}{1,0,0,0}
 \definecolor{MAGENTA}{cmyk}{0,1,0,0}
 \definecolor{YELLOW}{cmyk}{0,0,1,0}
}
\makeatother
\usepackage{babel}

\begin{document}

\title{Material slow and fast light in a zero-dispersion configuration}

\author{Bruno MACKE} 
\author{Bernard S\'{E}GARD}
\email{bernard.segard@univ-lille.fr}

\affiliation{Universit\'{e} de Lille, CNRS, UMR 8523, Physique des Lasers, Atomes et Mol\'{e}cules, F-59000 Lille, France}

 \date{ \today}

\begin{abstract}

We study the propagation of light pulses in an absorbing medium when the frequency of their carrier coincides with a zero of the refractive
index dispersion. Although slow light and,\emph{ a fortiori}, fast light are not expected in such conditions, we show that both can be
obtained by selecting particular phase-components of the transmitted field. Analytical expressions of the resulting signals are obtained
by a procedure of periodic continuation of the incident pulse and a proof of principle of the predicted phenomena is performed by means
of a very simple electrical network, the transfer function of which mimics that of the medium. 

\end{abstract}

%\setboolean{displaycopyright}{true}

%\begin{document}

\maketitle

\section{Introduction}\label{sec:Introduction}

The one-dimensional propagation of coherent light pulses with a \emph{slowly
varying envelope} through a \emph{linear}  medium is usually analyzed by means
of the group velocity \cite{bo02}. At the optical frequency $\omega$ it reads  $v_{g}\left(\omega\right)=c/\left[n(\omega)+\omega \left(dn/d\omega\right)\right]$ where $c$ is the light velocity in vacuum, $n(\omega)$ is the refractive index and its derivative $dn/d\omega$ is the  \emph{refractive index dispersion}. When the group velocity and the medium transmission are both constant over the whole pulse spectrum, the envelope travels undistorted at the velocity $v_{g}\left(\omega_{c}\right)=c/\left[n(\omega_{c})+\omega_{c} \left(dn/d\omega_{c}\right)\right]$  where $\omega_{c}$ is the pulse carrier frequency and $dn/d\omega_{c}$ is a short-hand notation of  $dn/d\omega$ for $\omega=\omega_{c} $ . When the above-mentioned double condition is not fulfilled, group velocity dispersion and transmission variation over the pulse spectrum result in a pulse reshaping. 

The refractive index dispersion  $dn/d\omega_{c}$ can take very large positive or negative values when the carrier frequency $\omega_{c}$ of the pulses is equal or close to the frequency $\omega_{0}$ of
a narrow and well-marked dip or peak of the medium transmission. When $dn/d\omega_{c}>0$
(normal dispersion), the group velocity is then very small compared
to the phase velocity $c/\left[n(\omega_{c})\right]$ (slow light
regime) while it becomes very large or even negative (fast last regime)
when $dn/d\omega_{c}<0$ (anomalous dispersion). The principle of
causality implies that the two regimes can be obtained with a same
medium, depending on the probe frequency $\omega_{c}$ \cite{bol93}.
We examine in the present article what occurs when $\omega_{c}$ is
such that $dn/d\omega_{c}=0$ (zero-dispersion configuration). Neither
slow light nor fast light are expected in this case. We will however
show that both can be observed by post-selecting particular phase
components of the transmitted field, in analogy with the experiments
involving post-selection of the field polarization \cite{so03,bru4,bi08,ma16}.
We specifically consider the reference case of a medium with a narrow
absorption line. Convincing demonstrations of slow light \cite{gri73,si09}
and fast light \cite{chu82,se85,ta03,ke12,je16} have been performed
with this system. The arrangement of our paper is as follows. In Section
\ref{sec:TRANSFER-FUNCTIONS}, we give the transfer functions for
the electric field and for its relevant phase components. The envelopes
of the corresponding transmitted pulses are determined in Section
\ref{sec:ENVELOPES} and we give in Section \ref{sec:EXPERIMENTS}
a proof of principle of the predicted phenomena by means of a very
simple electrical network. We conclude in Section \ref{sec:CONCLUSION}
by summarizing our main results.

\section{TRANSFER FUNCTIONS OF THE MEDIUM}\label{sec:TRANSFER-FUNCTIONS} 
Slow or fast light effects can be directly evidenced by comparing
the pulse transmitted by the medium (probe) to that which would be
transmitted in vacuum (reference). The transfer functions linking
the Fourier transforms of the corresponding fields to that of the
incident field read, respectively, $\exp\left[-i\tilde{n}(\omega)\omega\ell/c\right]$
and $\exp\left[-i\omega\ell/c\right]$, where $\tilde{n}(\omega)$
is the complex refractive index of the medium and $\ell$ its thickness
\cite{pa87}. In the time domain, the reference field is simply delayed
by the luminal transit time $\ell/c$ and this naturally leads to
use times retarded by $\ell/c$ to describe the transmitted fields.
In this \emph{retarded time picture} the transfer function for the
probe becomes 
\begin{equation}
H(\omega)=\exp\{-i[\tilde{n}(\omega)-1]\omega\ell/c\}\label{eq:zero}
\end{equation}
 and the reference field is equal to the incident field in real time.

\begin{figure}
\centering{}\includegraphics[width=0.95\columnwidth]{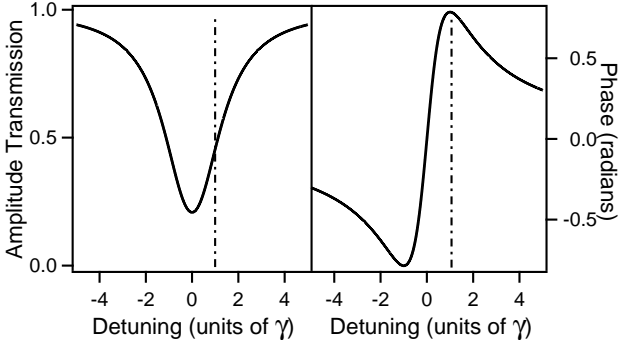} 
\caption{Modulus (on the left) and phase (on the right) of the field transfer
function vs. the optical detuning $\left(\omega-\omega_{0}\right)$
for $\alpha\ell=\pi/2$. The vertical dash-dotted line indicates the
detuning $\left(\omega_{c}-\omega_{0}\right)=\gamma$ of the pulse
carrier considered in the article.\label{fig:Figure1}}
\end{figure}

For the sake of simplicity, we consider a dilute medium with a Lorentzian absorption line of half width at half maximum $\gamma$
very small compared to the resonance frequency $\omega_{0}$ (narrow resonance limit). Under these conditions, the complex susceptibility $\chi(\omega)$ of the medium is such that $\left|\chi(\omega)\right|\ll1$  $\forall\omega$ and, in SI units, the complex refractivity $\widetilde{n}(\omega)-1$ is reduced to 
\begin{equation}
\widetilde{n}(\omega)-1=\sqrt{1+\chi(\omega)}-1\approx\chi(\omega)/2\label{eq:0a}
\end{equation}
In addition, the susceptibility is only significant when $\left|\omega\pm\omega_{0}\right|\ll\omega_{0}$.
For $\omega>0$, the classical Lorentz model \cite{bo02, ma96} and the semi-classical model of two-levels atoms \cite{al87,bo92} both lead to a relation of the form
\begin{equation}
\chi(\omega)\approx\frac{\gamma'}{\omega_{0}-\omega+i\gamma}\label{eq:0b}
\end{equation}
with $0<\gamma'\ll\gamma$. Eqs. (\ref{eq:zero}, \ref{eq:0a}, \ref{eq:0b}) finally yield
\begin{equation}
H(\omega)=\exp\left[-\frac{\alpha\ell}{1+i(\omega-\omega_{0})/\gamma}\right].\label{eq:un}
\end{equation}
where $\alpha=\omega_{0}\gamma'/(2 \gamma c)$ naturally appears as the medium absorption coefficient on resonance
\emph{for the amplitude}. By denoting $\Phi(\omega)$ the phase of $H(\omega)$, the group advance (the
opposite of the group delay) is given by the relation $a_{g}(\omega_{c})=d\Phi/d\omega\mid_{\omega=\omega_{c}}$ \cite{pa87}
and Eq.\ref{eq:un} yields
\begin{equation}
a_{g}(\omega_{c})=\left(\frac{\alpha\ell}{\gamma}\right)\left[\frac{1-\Delta^{2}/\gamma^{2}}{\left(1+\Delta^{2}/\gamma^{2}\right)^{2}}\right]\label{eq:deux}
\end{equation}
where $\Delta=(\omega_{c}-\omega_{0})$ is the detuning of the pulse
carrier frequency from resonance. The group advance attains its maximum
$a_{g}(\omega_{0})=\alpha\ell/\gamma$ for $\Delta=0$, is positive
(fast light regime) when $\left|\Delta\right|<\gamma$ and negative
(slow light regime) when $\left|\Delta\right|>\gamma$ . It cancels
when $\Delta=\pm\gamma$. We will consider in the following the case
where $\Delta=\gamma$ . Quite similar results are obtained when $\Delta=-\gamma$.
Figure \ref{fig:Figure1} shows the amplitude transmission $\left|H(\omega)\right|$
and phase $\Phi(\omega)$ as functions of the detuning $\omega-\omega_{0}$
in the reference case $\alpha\ell=\pi/2$. The vertical dash-dotted
line indicates the carrier frequency $\omega_{c}=\omega_{0}+\gamma$
considered in the following.

The transfer function $H_{\Delta}(\Omega)$ for the pulse envelope
is derived from Eq.(\ref{eq:un}) by passing in a frame rotating at
the frequency $\omega_{c} $\cite{al87}. We get
\begin{equation}
H_{\Delta}(\Omega)=\exp\left[-\frac{\alpha\ell}{1+i\left(\Omega+\Delta\right)/\gamma}\right]\label{eq:trois}
\end{equation}
where $\Omega=\omega-\omega_{c}$ with $\left|\Omega\right|\ll\omega_{c}$. On exact resonance ($\Delta=0$), $H_{0}\left(\Omega\right)=H_{0}^{*}\left(-\Omega\right)$, where the asterisk indicates complex conjugate. The corresponding impulse
response $h_{0}(t)$, inverse Fourier transform of $H_{0}\left(\Omega\right)$
\cite{pa87}, is then real \cite{cri70,ma18}. If the envelope $x(t)$
of the incident pulse is real (unchirped pulse) as assumed in the
following, the envelope $y(t)$ of the transmitted pulse will be also
real or, otherwise said, probe and reference fields will be in phase. In addition the relation $H_{0}\left(\Omega\right)=H_{0}^{*}\left(-\Omega\right)$ implies that the amplitude transmission $\left|H_{0}\left(\Omega\right)\right|$and
the phase $\Phi_{0}(\Omega)$ of $H_{0}\left(\Omega\right)$ are,
respectively, even and odd functions of $\Omega$. As discussed in
\cite{ma18}, the group advance $a_{g0}=(d\Phi_{0}/d\Omega)|_{\Omega=0}=\alpha\ell/\gamma$
can then be identified to the advance of the center-of-gravity (COG)
of $y(t)$ over that of $x(t)$ while $H_{0}(0)=e^{-\alpha\ell}$
is the ratio of the two envelopes areas \cite{ma18}. These results
hold whatever the pulse distortion is.

In the zero-dispersion configuration ($\Delta=\gamma$), $H_{\gamma}\left(\Omega\right)\neq H_{\gamma}^{*}\left(-\Omega\right)$
and the impulse response $h_{\gamma}(t)$ is complex. The envelope
$y(t)$ of the transmitted pulse is also complex. Its real part $y_{I}(t)$
and imaginary part $y_{Q}(t)$ are then the envelopes of the components
of the probe field, respectively, in phase (I) and in quadrature (Q)
with the reference field. The impulse responses $h_{I,Q}(t)$ linking
the envelopes $y_{I,Q}(t)$ of these two components to that of the
reference pulse read: 
\begin{equation}
h_{I}(t)=Re[h_{\gamma}(t)]=\frac{1}{2}\left[h_{\gamma}(t)+h_{\gamma}^{*}(t)\right]\label{eq:troisb}
\end{equation}
\begin{equation}
h_{Q}(t)=Im[h_{\gamma}(t)]=\frac{1}{2i}\left[h_{\gamma}(t)-h_{\gamma}^{*}(t)\right]\label{eqtroisc}
\end{equation}
Coming back in the frequency domain, we get the corresponding transfer
functions 
\begin{equation}
H_{I}\left(\Omega\right)=\frac{1}{2}\left[H_{\gamma}\left(\Omega\right)+H_{\gamma}^{*}\left(-\Omega\right)\right]\label{eq:quatre}
\end{equation}
\begin{equation}
H_{Q}\left(\Omega\right)=\frac{1}{2i}\left[H_{\gamma}\left(\Omega\right)-H_{\gamma}^{*}\left(-\Omega\right)\right]\label{eq:cinq}
\end{equation}
These transfer functions are closely related to those encountered
in Ref.\cite{ma16} where the post-selection was made on the field
polarization. As $H_{0}\left(\Omega\right)$ in the resonant case,
$H_{I,Q}\left(\Omega\right)=H_{I,Q}^{*}\left(-\Omega\right)$ and
the modulus $\left|H_{I,Q}\left(\Omega\right)\right|$ and phase $\Phi_{I,Q}(\Omega)$
are, respectively, even and odd functions of $\Omega$. As made in
the resonant case, we derive from Eqs.(\ref{eq:trois},\ref{eq:quatre},
\ref{eq:cinq}) the advance $a_{gI,gQ}$ of the COG of $y_{I,Q}(t)$ over that of $x(t)$ and the ratio $\left|H_{I,Q}(0)\right|$ of
its algebraic area over that of $x(t)$ . We get 
\begin{equation}
\gamma a_{gI}=\frac{\alpha\ell}{2}\tan\left(\frac{\alpha\ell}{2}\right)\label{eq:cinqa}
\end{equation}
\begin{equation}
H_{I}(0)=\cos\left(\frac{\alpha\ell}{2}\right)\exp\left(-\frac{\alpha\ell}{2}\right)\label{eq:cinqb}
\end{equation}
\begin{equation}
\gamma a_{gQ}=-\frac{\alpha\ell}{2}\cot\left(\frac{\alpha\ell}{2}\right)\label{eq:cinqc}
\end{equation}
\begin{equation}
H_{Q}(0)=\sin\left(\frac{\alpha\ell}{2}\right)\exp\left(-\frac{\alpha\ell}{2}\right).\label{eq:cinqd}
\end{equation}
The most significant results are obtained when $H_{I}\left(\Omega\right)$
and $H_{Q}\left(\Omega\right)$ are both minimum-phase-shift functions.
This requires that all the zeroes of their continuation in the complex
plane have a positive imaginary part \cite{pa87}. It is easily shown
that this condition is met when $\alpha\ell<\pi$. It then results
from Eqs.(\ref{eq:cinqa}, \ref{eq:cinqc}) that the COG of $y_{I}(t)$
is advanced (fast light regime) while that of $y_{Q}(t)$ is delayed
(slow light regime). We additionally remark that the transmissions
$\left|H_{I}(\Omega)\right|$ and $\left|H_{Q}(\Omega)\right|$ are,
respectively, minimal and maximal for $\Omega=0$.
\begin{figure}[h]
\centering{}\includegraphics[width=0.95\columnwidth]{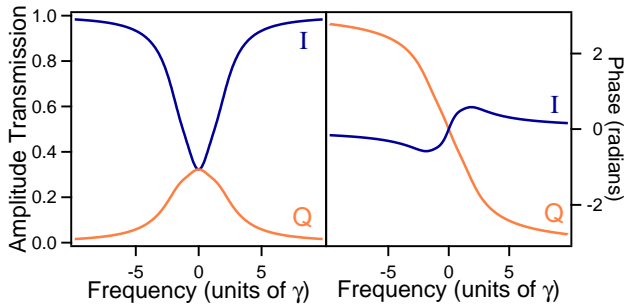} 
\caption{Same as Fig.\ref{fig:Figure1} for the envelope transfer functions
vs. the frequency $\Omega$. The blue (red) line refers to the in-phase
component I (the quadrature component Q). Note that, for the considered
optical thickness, the group advances $a_{gI}$ and $a_{gQ}$ (slope
of the corresponding phase at $\Omega=0$) have the same modulus but
opposite sign.\label{fig:Figure2}}
\end{figure}
Figure \ref{fig:Figure2} illustrates the previous points in the reference
case $\alpha\ell=\pi/2$ already considered in Fig.\ref{fig:Figure1}.
We are then in the remarkable situation where $a_{gQ}=-a_{gI}$ and
$H_{I}(0)=H_{Q}(0)$. This means that the COG of $y_{I}(t)$ and $y_{Q}(t)$
are shifted by the same amount in absolute value and that their \emph{algebraic}
area are equal.

A special attention will be paid in the following on fast light which
suffers from the most severe restrictions. Eq.(\ref{eq:cinqa}) might
lead to expect that advances as large as wanted could be obtained
for $\alpha\ell\rightarrow\pi$. It is worth remarking that this equation
only gives the advance of the pulse COG and that the corresponding
transmission $H_{I}(0)$ cancels. Anyway, obtaining large absolute
advances is not an aim \emph{per se} and the challenge in fast light
experiments is to attain ratios of the advances over the pulse duration
as large as possible with moderate distortion. The practical limitations
to these fractional advances are examined in the following section.

\section{ENVELOPES OF THE INCIDENT AND TRANSMITTED PULSES}\label{sec:ENVELOPES}

\begin{figure}[t]
\centering{}\includegraphics[width=0.95\columnwidth]{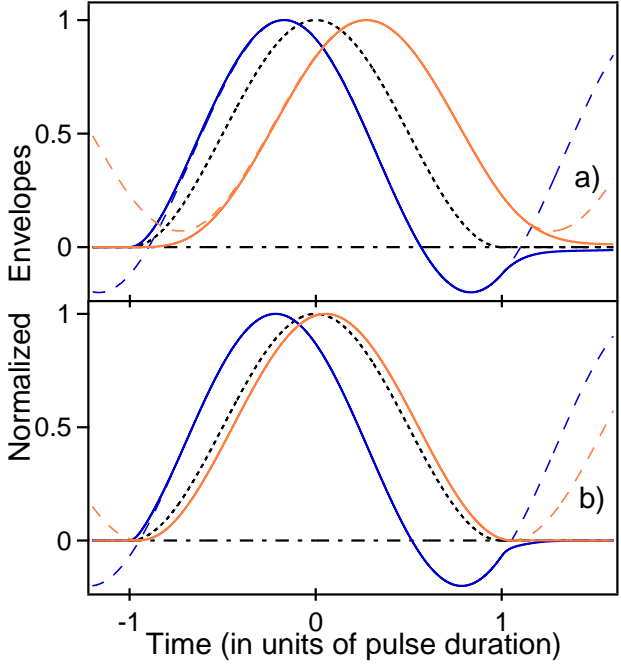} \caption{Normalized envelopes $y_{I}(t)$ (solid blue line) and $y_{Q}(t)$
(solid red line) vs. retarded time expressed in units of the incident pulse
duration $\tau=\pi/\Omega_{1}$. The dashed lines (same colours) are the solutions
obtained by periodically continuing the incident pulse. The envelope of the pulse transmitted in vacuum (reference pulse) is given for comparison (dotted black line). In real time all the envelopes should be shifted to the right by the luminal transit time $\ell/c$.
Parameters: $\alpha\ell$  ($\Omega_{1}/\gamma$) $=\pi/2$ ($1.17$) for a) and $3\pi/4$ ($0.345$) for b). \label{fig:Figure3}}
\end{figure}

We consider an incident pulse of carrier frequency $\omega_{c}=\omega_{0}+\gamma$
and of envelope 
\begin{equation}
x(t)=\cos^{2}\left(\frac{\pi t}{2\tau}\right)\Pi\left(\frac{t}{2\tau}\right)\label{eq:six}
\end{equation}
where $\Pi(u)$ designates the rectangle function equal to $1$ for
$-1/2<u<1/2$ and $0$ elsewhere. This pulse is very close to the
Gaussian pulse usually considered in the literature. It has a full
width at half maximum $\tau$ (taken as time unit in the following)
and a strictly finite overall duration $2\tau$. We exploit this point
by continuing the envelope $x(t)$ at every time by the periodic signal
\begin{equation}
\tilde{x}(t)=\cos^{2}\left(\frac{\pi t}{2\tau}\right)=\frac{1+\cos\left(\pi t/\tau\right)}{2}\label{eq:sept}
\end{equation}
$\tilde{x}(t)$ contains only three frequencies, namely $0$ and $\pm\Omega_{1}$
with $\Omega_{1}=\pi/\tau$. As shows Fig.\ref{fig:Figure3}, the
signals $\tilde{y}_{I,Q}(t)$ obtained by substituting $\tilde{x}(t)$
to $x(t)$ reproduce very well the main features of the exact envelopes
$y_{I,Q}(t)$ obtained by fast Fourier transform (FFT). We get: 
\begin{equation}
\tilde{y}_{I,Q}(t)=\frac{1}{2}\left\{ H_{I,Q}(0)+\left|H_{I,Q}(\Omega_{1})\right|\cos\left[\Omega_{1}t+\Phi_{I,Q}(\Omega_{1})\right]\right\} \label{eq:huit}
\end{equation}
$y_{I,Q}(t)$ has a maximum of amplitude $A_{I,Q}=\left[H_{I,Q}(0)+\left|H_{I,Q}(\Omega_{1})\right|\right]/2$
in advance over that of the reference pulse by $a_{I,Q}=\Phi_{I,Q}(\Omega_{1})/\Omega_{1}$.
The corresponding fractional advances read 
\begin{equation}
a_{I,Q}/\tau=\Phi_{I,Q}(\Omega_{1})/(\Omega_{1}\tau)=\Phi_{I,Q}(\Omega_{1})/\pi.\label{eq:neuf}
\end{equation}
The advances $a_{I,Q}$ of the maximum have the same sign that the
corresponding group advances $a_{gI,gQ}$ ($a_{I}>0$ , $a_{Q}<0$
). Since $H_{I}(0)<\left|H_{I}(\Omega_{1})\right|$, the amplitude
$A_{I}$ of the advanced signal is larger than its asymptotic value
$H_{I}(0)$ when $\Omega_{1}/\gamma\rightarrow0$ ($\gamma\tau\rightarrow\infty$),
the opposite occurring for the amplitude $A_{Q}$ of the delayed signal
{[}$A_{Q}<H_{Q}(0)${]}. Equation (\ref{eq:huit}) also enables us
to determine the full duration at half maximum $\tau_{I}$ and $\tau_{Q}$
of both signals. They read 
\begin{equation}
\tau_{I,Q}=\left(\frac{2\tau}{\pi}\right)\arccos\left[\frac{\left|H_{I,Q}(\Omega_{1})\right|-H_{I,Q}(0)}{2\left|H_{I,Q}(\Omega_{1})\right|}\right]\label{eq:dix}
\end{equation}
with $\tau_{I}<\tau$ (pulse narrowing) and $\tau_{Q}>\tau$ (pulse
broadening). All these results are consistent with those expected
for systems having a dip or a peak of transmission.

\begin{figure}
\begin{centering}
\includegraphics[width=0.95\columnwidth]{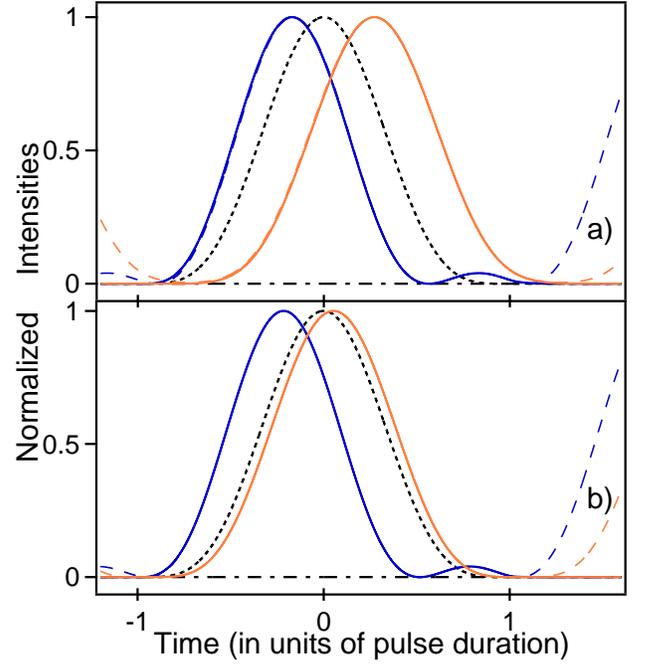}
\par\end{centering}
\caption{Same as Fig.\ref{fig:Figure3} for the intensity profiles $y_{I}^{2}(t)$
and $y_{Q}^{2}(t)$.\label{fig:Figure3bis}}
\end{figure}

In agreement with relativistic causality, both phase components start
at the same time $-\tau$ that the incident pulse in our retarded time picture  (Fig.\ref{fig:Figure3}).
As soon as it has a significant amplitude, the component $y_{Q}(t)$
is simply broadened and the (negative) advance $a_{Q}$ of its maximum
is generally close to the group advance $a_{gQ}$. The behaviour of
the in-phase component $y_{I}(t)$ is richer. The pulse distortion
of $y_{I}(t)$ is manifested in a narrowing (as already mentioned)
and, moreover, in the appearance of a secondary lobe \cite{ma18}.
The latter is also well reproduced by the periodic model (Fig.\ref{fig:Figure3}).
Quite generally, its maximum occurs at $t=\left[\pi-\Phi_{I}(\Omega_{1})\right]/\Omega_{1}$ and
its relative amplitude compared to that of the main lobe reads 
\begin{equation}
D=\frac{\left|H_{I}(\Omega_{1})\right|-H_{I}(0)}{\left|H_{I}(\Omega_{1})\right|+H_{I}(0)}\label{eq:douze}
\end{equation}
When it is small compared to unity, $D$ constitutes a good indicator
of the pulse distortion. The corresponding duration $\tau_{I}$ and
peak-amplitude $A_{I}$ of $y_{I}(t)$ given by the periodic model
read: 
\begin{equation}
\tau_{I}=\frac{2\tau}{\pi}\arccos\left[\frac{D}{1+D}\right]\label{eq:treize}
\end{equation}
\begin{equation}
A_{I}=\frac{H_{I}(0)}{1-D}\label{eq:quatorze}
\end{equation}
The envelopes shown Fig.\ref{fig:Figure3}a and Fig.\ref{fig:Figure3}b
have been obtained, respectively, for $\alpha\ell=\pi/2$ (reference
case) and for $\alpha\ell=3\pi/4$. In both cases, the pulse duration
$\tau$ has been chosen such that $D=20\%$. As illustrated Fig.\ref{fig:Figure3bis},
the distortion of the corresponding intensity profiles (currently
considered in optics) is quite moderate and the fractional advances
are $\sqrt{2}$ times larger than those of the envelopes. We note
in particular that the advanced intensity profiles compare favourably
with those observed in the reference experiments involving a gain-doublet
medium \cite{ste03}.

\begin{figure}[t]
\centering{}\includegraphics[width=0.95\columnwidth]{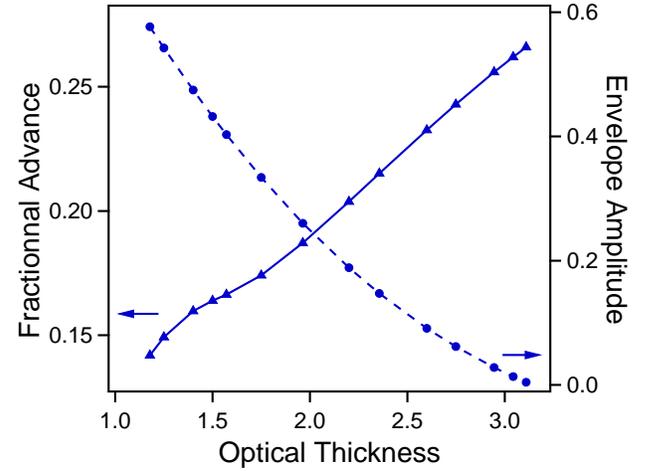}\caption{Fractional advance $a_{I}/\tau$ (solid blue line, left scale) and
amplitude $A_{I}$ (dashed blue line, right scale) of the maximum
of $y_{I}(t)$ vs. the optical thickness $\alpha\ell$ for $D=20\,\%$.\label{fig:Figure4}}
\end{figure}

\begin{figure}
\centering{}\includegraphics[width=0.95\columnwidth]{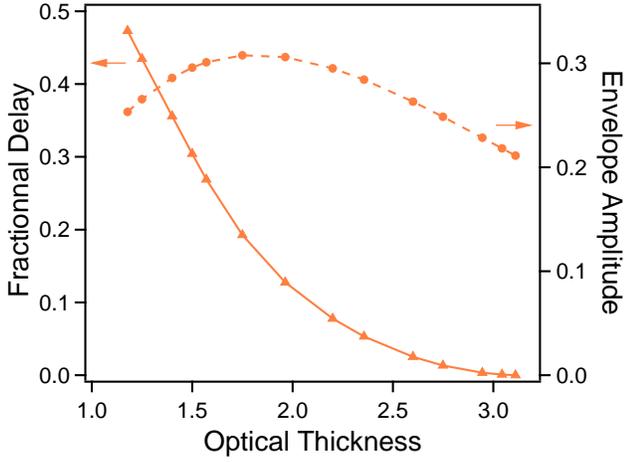}\caption{Same as Fig.\ref{fig:Figure4} for the fractional delay $-a_{Q}/\tau$
(solid red line, left scale) and amplitude $A_{Q}$ (dashed red line,
right scale) of the maximum of $y_{Q}(t)$.\label{fig:Figure5}}
\end{figure}

For a given distortion, relativistic causality imposes severe limitations
to the fractional advance $a_{I}/\tau$ of $y_{I}(t)$. As for every
fast light system, the larger is the dynamics of the system transmission,
the larger is the fractional advance \cite{ma05}. In the present
case, the transmission dynamics {[}maximum over minimum of $\left|H_{I}(\Omega)\right|${]}
is reduced to $1/H_{I}(0)$. Eq.(\ref{eq:quatorze}) thus involves
that the amplitude $A_{I}$ and the fractional delay $a_{I}/\tau$
evolve in opposite directions as functions of the optical thickness
$\alpha\ell$. Figure \ref{fig:Figure4} shows the results obtained
for $D=20\,\%$ in a broad domain of variation of $\alpha\ell$. In
the reference case $\alpha\ell=\pi/2$ (conditions of Fig.\ref{fig:Figure3}a),
we get $a_{I}/\tau\approx17\%$ and $A_{I}\approx0.40$ while $a_{I}/\tau\approx22\%$
and $A_{I}\approx0.15$ when $\alpha\ell=3\pi/4$ (conditions of Fig.\ref{fig:Figure3}b).
In the latter case, sufficient amplitude is conciliated with a fractional
advance which is not far below its asymptotic value ($\approx27\%$, see below). 
For the sake of completeness, we give Fig.\ref{fig:Figure5} the results
obtained in the conditions of Fig.\ref{fig:Figure4} for the delayed
envelope $y_{Q}(t)$. The variations of its amplitude $A_{Q}$ are
moderate, while the fractional delay $-a_{Q}/\tau$ continuously decreases
to $0$ when $\alpha\ell\rightarrow\pi$. We additionally mention
that, correlatively, the pulse duration $\tau_{Q}$ (the ratio $a_{Q}/a_{gQ}$)
regularly decreases (increases) to $\tau$ ( to $1$). 

When $\alpha\ell\rightarrow\pi$, the pulse duration required to obtain
a fixed distortion $D$ becomes very long. Denoting $\varepsilon=(\pi-\alpha\ell)/\pi$,
we get at the leading order in $\varepsilon$
\begin{equation}
\frac{\Omega_{1}}{\gamma}=\frac{\pi}{\gamma\tau}=\beta\epsilon\label{eq:quinze}
\end{equation}
where $\beta=2\sqrt{D}/\left(1-D\right)$. Putting this value in the
transfer functions $H_{I,Q}(\Omega_{1})$, we get, always at the leading
order in $\varepsilon$, the following asymptotic expressions of the
pulse advances and amplitudes 
\begin{equation}
\frac{a_{I}}{\tau}=\frac{1}{\pi}\arctan(\beta)\label{eq:seize}
\end{equation}
\begin{equation}
\frac{a_{gI}}{\tau}=\frac{\beta}{\pi}\label{eq:seizea}
\end{equation}

\begin{equation}
A_{I}=\frac{\pi e^{-\pi/2}}{4}\left(1+\sqrt{1+\beta^{2}}\right)\varepsilon\label{eq:dixsept}
\end{equation}
\begin{equation}
\frac{a_{Q}}{\tau}=\frac{a_{gQ}}{\tau}=-\frac{\beta\pi\epsilon^{2}}{4}\label{eq:dixhuit}
\end{equation}
\begin{equation}
A_{Q}\approx e^{-\pi/2}\label{eq:dixneuf}
\end{equation}
Eqs.(\ref{eq:seize}, \ref{eq:seizea}) shows that, strictly speaking,
$a_{I}\rightarrow a_{gI}$ only when $D\ll1$ but the fractional advance
$a_{I}/\tau$ is then very small. As for each fast light system, significant
fractional advances are paid by some distortion. The previous results
are quite consistent with those shown Figs.(\ref{fig:Figure4}, \ref{fig:Figure5})
for $D=20\,\%$, that is $\beta=\sqrt{5}/2$. For $\alpha\ell\rightarrow\pi$,
we get in particular $a_{I}/\tau\rightarrow27\%$, $A_{I}\rightarrow0$,
$a_{Q}/\tau\rightarrow0$, and $A_{Q}\rightarrow e^{-\pi/2}\approx0.21$.

\section{EXPERIMENTS WITH AN ELECTRICAL NETWORK}\label{sec:EXPERIMENTS}

In the optical experiments, the information on the phase shift induced
by the medium can be obtained by means of a frequency change translating
both reference and probe fields in the radiofrequency domain \cite{che05,lee13}.
The experiments are greatly simplified by working directly in this
domain. As back as 1961, Rupprecht \cite{ru61} evidenced
significant advances of the \emph{envelope} of the pulse transmitted
by a radiofrequency network with negative group delay (NGD). More
recent demonstrations of advanced pulse-envelope can be found in \cite{mi98,ca03,si04,ra13,ra14}.
However, as far as we know, the idea of phase post-selection to evidence
NGD effects is absent in all these experiments. 

\begin{figure}
\centering{}\includegraphics[width=0.95\columnwidth]{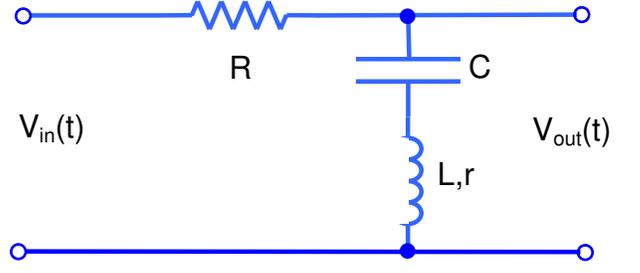}\caption{Electrical network used in our experiments. $r$ designates the resistance
$r_{L}$ of the inductor plus an eventual additional resistance $r_{a}$.
Indicative values of the parameters: $L=2.17\,mH$ , $C=153\,pF$
, quality factor of the inductor $Q=O(100)$ at $\omega=1/\sqrt{LC}$,
and $r_{a}=0$ or $33\,\Omega$, $R=1008\,\Omega$ or $365\,\Omega$
.\label{fig:Figure6 }}
\end{figure}

\begin{figure}[t]
\centering{}\includegraphics[width=0.95\columnwidth]{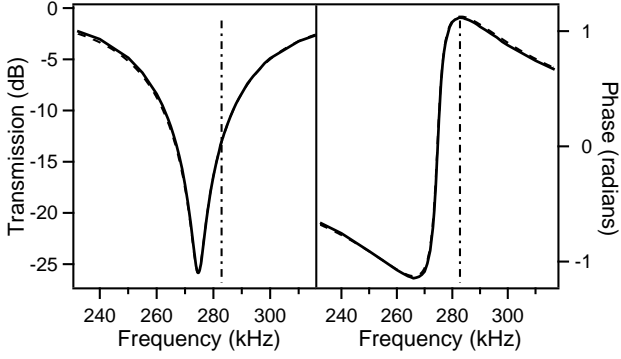}\caption{Transmission in dB (on the left) and phase (on the right) of the signal transfer function vs $\omega/2\pi$ for $r_{a}=0$ and $R=1008\,\Omega$
(solid line) and corresponding theoretical results for $\eta=0.226$
, $\omega_{0}/2\pi=274.9\,kHz$ and $\gamma/2\pi=8.53\,kHz$ (dashed
line). The vertical dash-dotted line indicates the zero-dispersion
frequency $\omega_{0}+\gamma$ .\label{fig:Figure7}}
\end{figure}

Figure \ref{fig:Figure6 } shows the very simple four port network employed in our experiments.
As the absorbing medium, it is purely passive. In the narrow resonance
limit, the transfer function relating the Fourier transform of the
output signal $V_{out}(t)$ to that of the input signal $V_{in}(t)$
reads 
\begin{equation}
H(\omega)=\frac{\eta+i\left(\omega-\omega_{0}\right)/\gamma}{\frac{1}{\eta}+i\left(\omega-\omega_{0}\right)/\gamma}\label{eq:vingtetun}
\end{equation}
where $\eta=\sqrt{r/(r+R)}$ ($0<\eta<1$), $\omega_{0}=1/\sqrt{LC}$
and $\gamma=\sqrt{r(R+r)}/(2L)$ with $\gamma\ll\omega_{0}$. The
general relation $a_{g}(\omega_{c})=d\Phi/d\omega\mid_{\omega=\omega_{c}}$
giving the group advance yields 
\begin{equation}
a_{g}(\omega_{c})=\frac{\left[(1/\eta)-\eta\right]\left(1-\Delta^{2}/\gamma^{2}\right)}{\gamma\left[1+\Delta^{2}/(\eta\gamma)^{2}\right]\left[1+(\eta\Delta)^{2}/\gamma^{2}\right]}\label{eq:vingtdeux}
\end{equation}
where $\Delta=\omega_{c}-\omega_{0}$ is the detuning of the pulse
carrier frequency from resonance. As for the absorbing medium, the
group advance is positive when $\left|\Delta\right|<\gamma$, is negative
when $\left|\Delta\right|>\gamma$, cancels when $\left|\Delta\right|=\gamma$
and is maximum on resonance where it takes the simple form $a_{g}(\omega_{0})=\left[\left(1/\eta\right)-\eta\right]/\gamma$.
The experimental transmission and phase obtained for $\eta=0.226$,
$\omega_{0}/2\pi=274.9\,kHz$ and $\gamma/2\pi=8.53\,kHz$ are shown
Fig.\ref{fig:Figure7}. 
They are in excellent agreement with those derived from Eq.(\ref{eq:vingtetun}).
Note, however, that the values of $\eta$, $\omega_{0}$ and $\gamma$
somewhat differ from those given below Eq.(\ref{eq:vingtetun}) which
are obtained by considering ideal components without including the
self-resonant behaviour of the capacitor and inductor \cite{si04}.

\begin{figure}
\centering{}\includegraphics[width=0.95\columnwidth]{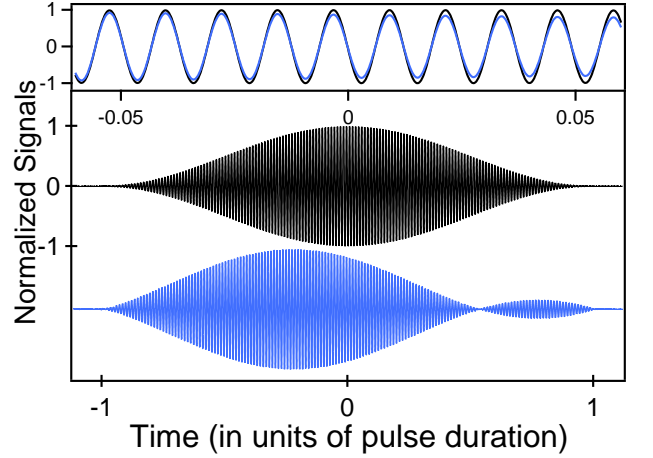} \caption{Normalized output signal $V_{out}(t)$ (blue line) vs. time expressed in units of the incident pulse duration $\tau$. This signal is obtained
in the resonant case ($\omega_{c}=\omega_{0}$). The input signal
$V_{in}(t)$ (black line) is given for reference. Parameters as in
Fig.\ref{fig:Figure7} and $\tau=295\mu s$ ($\gamma\tau=15.8$).
The advance of the maximum of $V_{out}(t)$ over that of $V_{in}(t)$
is $a=0.221\, \tau$ (corresponding group advance $a_{g}=0.266\, \tau$
) while its relative amplitude is $A=0.052$. Inset: comparison of
the two signals in the vicinity of $t=0$ showing that they are in
phase.\label{fig:Figure8}}
\end{figure}

\begin{figure}
\centering{}\includegraphics[width=0.95\columnwidth]{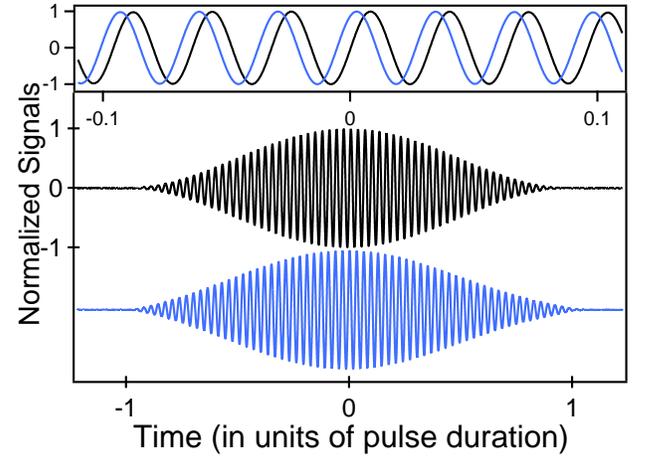}\caption{Same as Fig.\ref{fig:Figure8} in the zero-dispersion configuration
( $\omega_{c}=\omega_{0}+\gamma$). The output signal has a phase
differing from that of the input signal and the advance $a$ of its
maximum is negligible. The pulse duration $\tau=110\,\mu s$
has been chosen to facilitate the comparison with the results obtained
in the absorbing medium for $D=20\,\%$. It leads to $A=0.224$.\label{fig:Figure9}}
\end{figure}

In the time-resolved experiments, we use a waveform generator (Agilent
33500B) delivering both the sinewave signal of frequency $\omega_{c}$
and the modulation signal. It is used in the burst mode (single-shot
experiment). The signals $V_{in}(t)$ and $V_{out}(t)$ are sent on
two channels of a numerical oscilloscope (Keysight InfiniiVision DSOX4024A)
and both are acquired on 16000 points with a 10 bit vertical resolution.
Figure \ref{fig:Figure8} gives an example of signals obtained in
the resonant case ($\omega_{c}=\omega_{0}$) with the parameters of
Fig.\ref{fig:Figure7}. As expected, the maximum of $V_{out}(t)$
is significantly in advance over that of $V_{in}(t)$ but the two
signals are in phase. 
On the other hand, Fig.\ref{fig:Figure9}, obtained in the zero-dispersion
configuration ( $\omega_{c}=\omega_{0}+\gamma$), confirms that the
advance is then negligible but that the two signals are not in phase.

The transfer functions $H_{\gamma}(\Omega)$, $H_{I}(\Omega)$ and
$H_{Q}(\Omega)$ for the envelopes in the zero-dispersion configuration
($\Delta=\gamma$) are derived from $H(\omega)$ as those of the absorbing
medium. For the electrical network $H_{\gamma}(\Omega)$ reads 
\begin{equation}
H_{\gamma}(\Omega)=\frac{\eta+i+i\Omega/\gamma}{1/\eta+i+i\Omega/\gamma}.\label{eq:vingttrois}
\end{equation}
As in the absorbing medium case, the transfer functions $H_{I}(\Omega)$
and $H_{Q}(\Omega)$ are deduced from $H_{\gamma}(\Omega)$ by Eqs.(\ref{eq:quatre},\ref{eq:cinq}).
They yield in particular 
\begin{equation}
H_{I}(0)=\frac{2\eta^{2}}{1+\eta^{2}}\label{eq:vingtquatre}
\end{equation}
\begin{equation}
\gamma a_{gI}=\frac{\left(1-\eta^{2}\right)^{2}}{2\eta\left(1+\eta^{2}\right)}\label{eq:vingtcinq}
\end{equation}
\begin{equation}
H_{Q}(0)=\frac{\eta(1-\eta^{2})}{1+\eta^{2}}\label{eq:vingtsix}
\end{equation}
\begin{equation}
\gamma a_{gQ}=-\frac{2\eta}{\left(1+\eta^{2}\right)}\label{eq:vingtsept}
\end{equation}
For a given distortion $D\ll1$, the upper limit of the fractional
advance $a_{I}/\tau$ is again approached when $H_{I}(0)\rightarrow0$,
that is when the pulse amplitude $A_{I}\rightarrow0$. By calculations
quite similar to those leading to Eqs.(\ref{eq:quinze}-\ref{eq:dixneuf}),
we retrieve the upper limit $a_{I}/\tau=\left(\arctan\beta\right)/\pi$
($27\%$ for $D=20\%$) obtained in the absorbing medium case.

The post-selection of the in-phase and quadrature components of the
output signal $V_{out}(t)$ is experimentally performed as follows.
The data collected by the numerical oscilloscope are treated by computer.
In a first step, we generate a continuous sinewave, the frequency
and \emph{phase} of which coincide with those of the input signal
$V_{in}(t)$. 
\begin{figure}[t]
\centering{}\includegraphics[width=0.95\columnwidth]{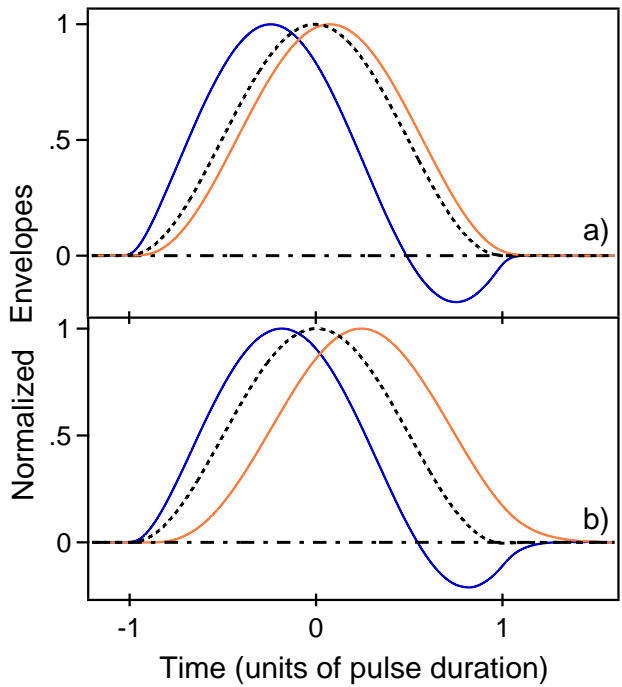}\caption{Two examples of envelopes $y_{I}(t)$ (solid blue line) and $y_{Q}(t)$ (solid red line) experimentally obtained in the zero-dispersion configuration.
The envelope $x(t)$ of the input signal (dotted black line) is given
for reference. (a) is obtained in the conditions of Fig.\ref{fig:Figure9},
viz. $\tau=110\,\mu s$, $\eta=0.226$ and $\gamma/(2\pi)=8.53\,kHz$.
We get in this case $a_{I}=0.24\,\tau$, $A_{I}=0.12$ for the in-phase
component and $a_{Q}=-0.074\,\tau$, $A_{Q}=0.20$ for the quadrature
component. (b) is obtained for $\tau=76\,\mu s$ by using
a circuit with resistances $R=365\,\Omega$ and $r_{a}=33\,\Omega$
leading to $\eta=0.45$ and $\gamma/(2\pi)=7.45\,kHz$. We have in
this case $a_{I}=0.20\,\tau$, $a_{Q}=-0.24\,\tau$ and $A_{I}/A_{Q}=1.37$.\label{fig:Figure10}}
\end{figure}
This continuous sinewave is next multiplied by the output
signal $V_{out}(t)$ to deliver the envelope $y_{I}(t)$ of the in-phase
component (I), the harmonic at $2\omega_{c}$ and the high frequency
noise being eliminated by a finite impulse response (F.I.R.) filter.
The used low pass filter (IGOR software Blackman 367) insures a rejection
better than 70 dB for $\omega/(2\pi)>200\,kHz$. The envelope $y_{Q}(t)$
of the quadrature component is similarly derived by using a continuous
sinewave in quadrature with that used to obtain $y_{I}(t)$ .

Figure \ref{fig:Figure10} shows the envelopes $y_{I}(t)$ and $y_{Q}(t)$
experimentally observed in the zero-dispersion configuration ($\Delta=\gamma$)
in two representative cases. As it was made to obtain Fig.\ref{fig:Figure9}
also as Figs.(\ref{fig:Figure3}-\ref{fig:Figure5}) for the absorbing
medium, the durations $\tau$ of the incident pulse are chosen such
that $D=20\,\%$ for the I-component. The envelopes shown Fig.\ref{fig:Figure10}a
are observed in the conditions of Fig.\ref{fig:Figure9}. They are quite comparable to those obtained theoretically with an absorbing
medium of optical thickness $\alpha\ell=3\pi/4$ (see Fig.\ref{fig:Figure3}b).
In particular, the advance $a_{I}$ is significantly larger than the
delay $-a_{Q}$. On another hand, the amplitudes of the two components
are such that $A_{Q}/A_{I}\approx1.7$ and this explains why no secondary
lobe is visible Fig.\ref{fig:Figure9} in the overall envelope $y(t)=\sqrt{y_{I}^{2}(t)+y_{Q}^{2}(t)}$.
As a second example, Fig.\ref{fig:Figure10}b shows the envelopes
experimentally observed when $\eta=0.45$ and $\gamma/(2\pi)=7.45\,kHz$.
In this case $\left|a_{I}/a_{Q}\right|$ and $A_{Q}/A_{I}$ are both
close to unity and the envelopes are now comparable to those obtained
in the case $\alpha\ell=\pi/2$ taken as reference for the absorbing
medium (see Fig.\ref{fig:Figure3}a). In case a) as in case b), the
observed envelopes are in very good agreement with the envelopes derived
by FFT by using the transfer function $H_{\gamma}(\Omega)$ given
Eq.(\ref{eq:vingttrois}). In addition, the advances, amplitudes and
pulse durations are exactly determined by the periodic model with,
in particular, $\tau_{I}=0.89\,\tau$ and $A_{I}=1.25\,H_{I}(0)$ as
predicted by Eqs.(\ref{eq:treize},\ref{eq:quatorze}).

\section{CONCLUSION}\label{sec:CONCLUSION}

The dilute medium with a narrow absorption line is a reference system
for the observation of fast and slow light. Fast light is obtained
when the carrier frequency of the incident pulse coincides or is close
to resonance while slow light is observed when this frequency lies
in the line wings. There are thus two intermediate carrier frequencies
for which the group velocity equals that of the light in vacuum. Paradoxically
enough, we have shown that, in such a case, fast and slow light can
be simultaneously observed. This is achieved by post-selecting particular
phase components of the transmitted field. Fast light is obtained
by selecting the component in phase with that of a pulse travelling
the same distance in vacuum while slow light is observed on the quadrature
component. The general properties of fast and slow light are retrieved
with this arrangement. A particular attention is paid to fast light
to which the relativistic causality imposes the most severe constraints.
As usual, evidencing significant fast light effects with moderate
distortion requires large transmission dynamics of the medium and
long incident pulses. Finally the theoretical results obtained in
optics with an absorbing medium are experimentally reproduced by using
a passive electrical network running in the radiofrequency range.
We expect that our work will stimulate direct demonstrations in optics
or microwave. In this purpose, we emphasize that the phase post-selection
procedure introduced in the present article can be applied to different
frequency configurations and systems.

\large\textbf{Funding}: \normalsize
Contrat de Plan Etat-Région (CPER), Photonics and Society Project (P4s) ; 
Agence Nationale de Recherche (ANR), LABEX CEMPI Project (ANR-11-LABX-0007).

\large\textbf{Disclosures}:\normalsize  The authors declare no conflicts of interest.

\end{document}